# Fracture simulation for zirconia toughened alumina microstructure

**Kyungmok Kim(1), Jean Geringer(2), Bernard Forest(3)**

(1) SCHOOL OF AEROSPACE AND MECHANICAL ENGINEERING, KOREA AEROSPACE UNIVERSITY

(2) CENTRE CIS, DEPARTEMENT STBIO, IFRESIS IFR 143, LGF UMR 5307

(3) CENTRE CIS, DEPARTEMENT STBIO, IFRESIS IFR 143, LCG UMR 5146

## 1. Introduction

Alumina and zirconia have improved the behaviour of artificial articular joints such as hip prostheses. In addition, their chemical inertness and good tribological properties give them an advantage over metals or polymers for a bearing component such as a head of hip prosthesis. However, these materials maintain drawbacks such as the body reaction around the joints. Wear debris from the materials induces bone resorption, consequent inflammatory reaction and implant loosening. Bone resorption and aseptic loosening are the major problems limiting the lifetime of current implants. Methods improving durability of orthopaedic implants are being explored by proposing new materials. New materials include zirconia toughened alumina (ZTA) microstructured composites. ZTA composites maintain numerous advantages such as high mechanical properties of yttria-stabilized zirconia, high inertness of alumina without any of their respective drawbacks, high hardness for improved wear behaviour, high toughness and crack propagation threshold for higher reliability, and high chemical inertness for longer durability. These properties will allow the design of more reliable and long-lasting implants for less invasive surgery.

In hip prosthesis, material removal occurs resulting from repeated mechanical shocks [1].

Dispersion of stabilized $ZrO_2$ particles in zirconia toughened alumina composites should lead to the improvement of shock resistance in hip prosthesis. Manufacturing processes for ZTA composites include sintering for ceramic densification [2].

A typical sintering process for ZTA composites consists of heating from room temperature to 1500 °C, then a thermal treatment of 1500 °C for 2 hours, and cooling up to room temperature. The mismatch of thermal expansion coefficient between $Al_2O_3$ and $ZrO_2$ grains induces residual stresses in the microstructure of ZTA composite. Particularly, most residual stress is arisen during cooling, directly affecting fracture behaviour under loading conditions.

Several experiments were conducted in order to investigate failure of hip prosthesis [3-6]. Shock experiments were performed for measuring wear rate of alumina heads or cups of hip prosthesis [3,4]. The surfaces of alumina heads were progressively locally worn out with respect to number of cycles. Wear and fracture of zirconia heads against alumina cups were investigated with a hip simulator [5]. Wear rate between the head and the cup was measured under micro-separation conditions. Crack growth resistance of alumina-zirconia ceramics were experimentally determined [6]. Crack growth rate of alumina toughened zirconia was lower than those of other ceramic biomaterials such as pure alumina or zirconia.

A variety of methods for investigating failure behaviour are used including fracture mechanics, continuum damage mechanics, and so on. Particularly, a method using a cohesive zone law is useful for fracture behaviour of micro-and macro-structures, since this method enables simulating fracture of interfaces between physical parts and characterizing post-yield softening. Elements used in cohesive zone modelling do not represent any physical material but contain cohesive forces arising when they are pulled apart. If these elements satisfy a pre-defined damage criterion, cohesive forces are removed and no stresses remain between physical parts.

This traction-separation response in a cohesive zone law can be expressed with bilinear, exponential, power-law, polynomial, or trapezoidal forms. A cohesive zone model was used for assessing for simulating and predicting the onset and the propagation of the delamination in a composite [7]. A numerical tool using a cohesive zone law was developed and was applied to finite element method. Some cohesive models were developed for describing fracture of parts of the human body [8-10]. Tension and compression cracks in bone-cement interface were investigated by using cohesive zones [8]. A cohesive fracture model was formulated, especially for human femoral cortical bones [9]. A micro-mechanical model using cohesive zone theory was developed [10]. Mechanical damages in the ligament-to-bone attachment of a human knee joint were predicted. The cohesive zone model theory was focused on the development of behaviour laws for crack initiation and propagation at an interface within a fibrous material or at the interface between materials. The role of cohesive properties on intergranular crack propagation in brittle polycrystals was studied [11]. The relation between crack path and cohesive law parameters was investigated. Random realizations of a polycrystalline topology were considered instead of using actual microstructures.

Numerical studies for investigating fracture of ceramic microstructures were carried out using finite element methods. A two-dimensional finite element model of compressive fracture in ceramics was developed [12]. Finite element modelling was performed with pure alumina microstructures generated by Voronoi tessellation of randomly positioned seeds. Micro- and macro-scale boundary element modelling was performed with polycrystalline $Al_2O_3$ ceramics [13]. Three-point bending was applied to a model for inducing crack propagation. A two-dimensional finite element model was developed for microstructures of pure $Al_2O_3$ used for human hip prostheses [14]. Image-based extended finite element modelling was employed for thermal barrier coatings [15]. Two-dimensional XFEM analysis of actual microstructure containing cracks was developed under assumption of plane stress condition.

However, a finite element model for zirconia toughened alumina composites has not been developed despite its importance. Meanwhile, it may be possible to induce strain singularities close to triple junctions in a microstructure without fine meshing. An extended finite element method for polycrystals with discontinuous grain boundaries was proposed for avoiding singularities [16, 17]. For the purpose of avoiding singularity problem, special enrichment functions describing singular behaviour were implemented in extended finite element model. The developed finite element model using the enrichment function was used for predicting lifetime of solder joints [18] and for simulating fracture in electronic package applications [19].

In dense $Al_2O_3$-10 vol% $ZrO_2$ microstructures, $Al_2O_3$ and $ZrO_2$ grains vary in shape and size, thereby leading to difficulties in modelling. Image digitalization software and commercial finite element software (ABAQUS® 6.8) enables modelling such complex grains and grain boundaries. In this study, actual $Al_2O_3$ and $ZrO_2$ grains were modelled with a method developed by the authors. A two-dimensional finite element model was developed using a cohesive zone law. In this study, fine meshing was employed to avoid strain singularities that may occur close to the triple junctions in an actual microstructure. Fracture and fatigue behaviours of a microstructure were then investigated. Magnitudes of imposed stresses were higher than those found at the microscopic scale. Body fluid exists in actual hip prostheses, constituted by proteins (Ex. globulin, albumin), salts as chloride sodium, chloride calcium, and so on [20]. But, in this study, dry contact condition without the body fluid was considered between a femoral head and an acetabular cup; dry contact could occur after disruption of the lubrication film during human gait.

## 2. Finite element simulation

A bilinear, time-independent cohesive zone law was used for describing fracture of grain boundaries in this study. The cohesive zone law with traction-separation response allows specification of mechanical properties such as stiffness, strength, and relative displacement at failure [21]. The bilinear cohesive response is determined with maximum nominal stresses (normal and shear) and displacements (normal and shear) at failure. Once the maximum displacement of a cohesive element exceeds pre-described failure magnitude, stresses of a cohesive element become fully released. The bilinear cohesive zone law was illustrated in Fig 1 (right) [14].

[Fig. 1]

Fig. 1(left) presents a proposed simulation algorithm for determining damage of a microstructure. A two-dimensional finite element model including grains and cohesive zones is generated by digitalizing an actual microstructure image. Material properties of grains and cohesive zones are then defined. Loading and boundary conditions similar to those found at hip prosthesis are applied to models. Plane strain, implicit analysis is performed in the simulation. During simulation, stresses and strains of grain elements are calculated along with damage variables of cohesive elements. Fracture of grain boundaries is evaluated in terms of the damage variable. When the damage variable of an element is equal to unity, the element is deleted and the deleted element is considered as crack.

Commercial finite element software (ABAQUS® 6.8 standard) was used for simulating models. Each model was simulated on a cluster with 10 calculation nodes (Intel Xeon Quad Core 3 GHz, 64 bits with 1 GB). Viscous stabilization was implemented to avoid severe convergence arising from softening behaviour and stiffness degradation, as recommended by ABAQUS [21]. Dissipated energy fraction specified for viscous stabilization was 0.002~0.005 (0.2~0.5%) during calculation.

[Fig.2]

Fig. 2a and 2d show the microstructure images of $Al_2O_3$-10 vol% $ZrO_2$ ceramics [22]. Fig. 2b illustrates a generated model with Fig 2a, including 26 $Al_2O_3$ grains and 24 $ZrO_2$ ones. Dark bold lines in the model represent cohesive layers. Fig. 2e is the other model generated with Fig. 3d by the same modelling procedure. Fig. 2c and 2f show models with a void. A single alumina grain was deleted for investigating the effect of a void in a composite. The size of all models is 5 µm × 5 µm. All parts were meshed sufficiently finely for guaranteeing appropriate degree of resolution of local stress concentration effects (11000 elements for $Al_2O_3$ grain, 1163 elements for $ZrO_2$ grain, 1076 elements for the layer between $Al_2O_3$ grains, and 1247 elements for the layer between $Al_2O_3$ and $ZrO_2$). Particularly, the size of a cohesive element was 0.05 µm × 0.005 µm.

[Table 1]

[Fig. 3]

[Table 2]

In this study, conditions, found at the contact surface between a head and a cup, were taken into account. That is, the selected microstructures were assumed to be placed at contact surface.

Because initial surface roughness was about 0.02 µm, the surface roughness in the microstructures was ignored. Contact stresses between a head and a cup in hip prosthesis were reproduced by applying normal and shear stresses to the top surface of a model. The bottom surface was fixed in all directions while loading was being applied. The left and right sides of a model were not fixed, thus the model was permitted to undergo horizontal and vertical deformations.

Little is known about micro-mechanical properties of ZTA composite. Thus, the following assumptions were made in this study.

- The magnitude of normal stress varies according to location within an actual contact surface due to surface roughness, ranging from 5 MPa to 500 MPa [3]. In this study, four normal stress magnitudes of 50, 100, 250 and 500 MPa were selected for investigating the effect of contact stress.

- Dry contact condition and room temperature were considered. Coulomb friction coefficient of the material is about 0.4 under these conditions [23,24]. Shear stress was determined according to the magnitude of normal stress and friction coefficient (i.e. shear stress = 0.4 × normal stress).

- Grains behaved elastically as shown Table 1.

- The magnitude of elastic modulus was differently defined according to the direction. $E_x$ was ten percent lower than a typical value of elastic modulus of the material (374 MPa for $Al_2O_3$ and 210 MPa for $ZrO_2$), and $E_y$ was ten percent greater than the value.

- Poisson ratio values of $Al_2O_3$ and $ZrO_2$ grains were defined as 0.22 and 0.3, respectively.

- Each grain in $Al_2O_3$-10 vol% $ZrO_2$ microstructure maintained a differently assigned

material orientation, defined by an angle subtended from the horizontal axis as shown in Fig. 4. Each angle was chosen according to the shape of a grain. That is, x axis was placed in the longitudinal direction of a grain.

- The mechanical property of cohesive layers was different according to grain boundaries, since selected microstructures have two sorts of grain boundaries: $Al_2O_3$ vs. $Al_2O_3$ and $ZrO_2$ vs. $Al_2O_3$. Mechanical properties of cohesive zones are shown in Table 2.

- Fracture energy release rate of the layer between $Al_2O_3$ grains was equal to 0.5 J m$^{-2}$ in the pure tensile mode. This fracture energy release rate was uniformly distributed to all the cohesive layers between $Al_2O_3$ grains. The strength magnitude of a cohesive layer defined here is similar to those found in the literature [12,25]. Fracture energy release rate of the layer between $ZrO_2$ and $Al_2O_3$ grains was higher than the value between $Al_2O_3$ grains.

- The nominal shear stress was determined on the basis of the grain boundary shear strength's dependence on the compressive yield strength (3000 MPa) similar to friction coefficient (0.13) of a metallic glass [26].

## 3. Results

### 3.1. Stress distribution in a microstructure after cooling

A sintering process of $Al_2O_3$-10 vol% $ZrO_2$ leads to ceramic densification, resulting to the increase of mechanical strength. After a sintering process, residual stresses remain due to thermal expansion mismatch between $Al_2O_3$ and $ZrO_2$. In this simulation, linear thermal expansion coefficients of grains were defined as shown in Table 3.

[Table 3]

[Fig. 4]

After cooling a microstructure from 1500 °C to 25 °C, stresses of grains in a microstructure were determined. The magnitudes of stresses vary according to locations of the microstructure due to local stress concentration, ranging from -500 MPa to 700 MPa. Thus, the average stress value of grains was computed as Fig. 4. It was identified that $ZrO_2$ grains were subjected to tension while $Al_2O_3$ ones underwent compression in both models. Small shear stresses of both grains remained after cooling. The average magnitude of the compressive stress in $Al_2O_3$ grains was lower than the value (150 ± 50 MPa) found in the literature [27]. The alumina-zirconia composite considered in the literature contained 22 percent zirconia in volume, greater than the amount of zirconia in this study.

### 3.2. Crack distribution at various static loadings

A sintering process plays a role in minimising porosity by densificating a ceramic structure. However, voids may occur unpredictably after manufacturing processes. The existence of voids leads to the decrease of mechanical strength, thereby inducing micro-cracks. The effect of a void was investigated by modelling a microstructure containing a single void. Finite element models with a void and without a void were simulated under the same loading and boundary conditions. All models were sintered prior to being loaded.

[Fig. 5]

[Table 4]

Fig. 5 shows crack density versus applied normal stress chart. Crack density was defined as the ratio of total crack length to the total length of grain boundaries in a model. Here, the total crack length was defined as the sum of longitudinal distances of cracks generated in a microstructure. The total crack length was measured after a loading-unloading cycle. It was apparent that crack density values of microstructures with a single void were higher than those without a void. Models A and B were simulated under four different loading conditions. In both models, the crack density gradually increased with increasing the magnitude of applied stress within range of 50-500 MPa. A notable result was that the difference of crack density existed between two models though crack density rates were similar as shown in Table 4. This difference shows that total crack length is affected by grain shape and arrangement.

[Fig. 6]

[Fig. 7]

Fig. 6 illustrates crack distribution of model A under different loading conditions. At a normal stress of 100 MPa, a few small cracks (< 0.1 µm) occurred at the upper-left side of the model. At 500 MPa, long cracks were observed at the upper side of a model (Fig. 6b). The model with a void maintained more cracks, compared with that without a void (Fig. 6c). In addition, locations where the cracks occurred were different. At 500 MPa, long horizontal cracks were found at the model with a void, thereby making a grain separated from the model (Fig. 6d). Fig. 7 illustrates crack distribution of model B under two loading conditions. At a normal stress of 100 MPa, small cracks (< 0.2 µm) were observed at a variety of locations. At 500 MPa, long cracks were found at the upper side of the model, but grains were still adhered (Fig. 7b). Fig. 7c and Fig. 7d show that longer cracks were generated due to a void, compared with results without a void. These results show that crack length as well as crack path is somewhat different according to a microstructure.

### 3.3. Crack distribution after cyclic loadings

[Fig. 8]

[Fig. 9]

Fatigue behaviour of the sintered models was investigated by applying repeated contact stresses. During simulation, peak values of the normal stress and the shear stress were 100 MPa and 40 MPa, respectively. Fig. 8 illustrates crack distribution of model A with a void. After the initial fatigue cycle, small cracks occurred at the upper-right side of the model. These cracks grew with increasing number of fatigue cycles. Long cracks were then generated by combining small ones. After the tenth fatigue cycle, grain boundaries were seriously damaged as a result of progressive crack growth. A few grains were found to be separated. Fig. 9 shows crack distribution of model B with a void. It was also observed that small cracks occurred after the initial cycle and grew with respect to number of cycles. After the fifth cycle, most cohesive elements around an upper-right grain were removed. It is demonstrated that this proposed method allows simulating progressive crack growth by applying repeated contact stresses to a microstructure.

[Fig. 10]

Fig. 10 shows crack density evolution of the models. The slope of the plot corresponds to crack density rate. Each crack density rate can be expressed as a linear function of the number of cycles. The crack density rate of model B was higher than that of model A. In addition, it was found that a void led to the increase of crack density rate. This result shows that crack density rate varies according to a microstructure.

## 4. Discussion

Material loss rate in a model with a void can be approximately evaluated with the amount of separated grains with respect to number of cycles. If the amount of cohesive elements remaining between grains is so small that the grains can be considered to be separated in Fig. 8c, the loss area in model A is approximately $1.0 \times 10^{-6}$ mm$^2$ after ten cycles (the cross section of the material is 25 µm$^2$). Supposing that grain thickness is 0.001 mm, the loss rate is determined as $0.1 \times 10^{-9}$ mm$^3$/cycle, i.e. $10^{-4}$ mm$^3$/million of cycles for 25 µm$^2$. Meanwhile, it was identified from shock experiments that the actual cross section of a head submitted to shocks was approximately 5 µm x 3 mm [4]. Thus, the loss rate is equal to 1.5 mm$^3$/million of cycles. This wear rate is similar to the one measured by Stewart, i.e. 1.84 mm$^3$/million of cycles [5]. Meanwhile, in the authors' shock degradation experiment with ZTA composite, it was identified that wear volume is equal to 0.12 ± 0.01 mm$^3$/million of cycles [28]. From the microscopic to the macroscopic scale, our calculation is an order of magnitude. The calculated loss rate corresponds to a local wear rate at contact surfaces of hip prosthesis, since microstructures are used. For more accurate calculation of average wear volume, various actual microstructures found at the contact surface needs to be used. It might be necessary to compare calculated wear volume with clinical observations of zirconia toughened alumina. Recently, insertion of zirconia toughened alumina into patient's body is begun. Thus, clinical observation of zirconia toughened alumina microstructures is very limited. Zirconia toughened alumina for hip prosthesis is expensive compared with pure alumina. This proposed method is also useful in the design and manufacture process of materials. It is possible to perform direct comparison among microstructures in terms of fracture. This developed simulation could decrease the cost of zirconia toughened alumina.

In this simulation, dry contact condition was considered, thus full fluid film lubrication contact condition needs to be modelled. In addition, external stresses were applied to the models for reproducing deformation of a microstructure at the contact surface of a head. Displacement could be imposed to the models instead of the shear stress; this modelling can be achieved after identifying actual relative displacement between a head and a cup.

## 5. Conclusions

This paper describes finite element modelling for fracture and fatigue behaviours of zirconia toughened alumina (ZTA) ceramic at the micro level. ZTA ceramic is a promising material for hip prosthesis due to its superb mechanical properties. Nevertheless, modelling of fracture behaviour of the ceramic is rather limited. In this paper, a two-dimensional finite element model was developed for investigating crack initiation and propagation of the microstructure of ZTA ceramic. Two different $Al_2O_3$-10 vol% $ZrO_2$ microstructures were selected and modelled with commercial finite element software and an image digitalization program developed by the authors. Simulation algorithm from model generation to analysis was detailed, enabling cyclic loadings. A bilinear, time-independent cohesive zone law was implemented for describing fracture behaviour of grain boundaries. The following was carried out with this simulation.

First, residual stress modelling was performed with the microstructure. Residual stresses were determined arising from the mismatch of thermal expansion coefficient between alumina and zirconia grains. Results showed that compressive stresses remained in alumina grains and tensile stresses remained in zirconia ones.

Second, the effect of contact stresses in the microstructure was investigated in terms of the crack density. Normal stress and shear stress were applied to the top surface of the models. The total crack length in a model was then measured. It was identified from results that crack density was gradually increased with increasing the magnitude of the stress. In addition, void effect was identified by removing a single grain in a microstructure. It was observed that a void in a microstructure led to the increase of crack density.

Finally, fatigue behaviour of a microstructure was investigated by performing simulations under cyclic loadings. It was apparently observed that crack density linearly increased with respect to number of cycles. Meanwhile, difference of crack density rates existed between selected microstructures. It was also identified that a void led to the increase of crack density.

In conclusion, this proposed finite element simulation offers an effective method for modelling residual stress and for investigating fracture behaviour of ZTA microstructures. Further investigation will focus on three-dimensional finite element modelling and modelling and consideration of full fluid film lubrication condition.

**Figure captions**

**Fig. 1** Simulation algorithm for determining damage. $N_i$ denotes the $i^{th}$ cycle.

**Fig. 2** Images of $Al_2O_3$-10%$ZrO_2$ microstructures [14] and finite element models. The middle images (b, e) are models generated with the actual images (a, d). For convenience, (b) and (e) are denoted as model A and model B, respectively. A single alumina grain in each model was deleted for investigating the effect of a void (c, f).

**Fig. 3** Schematic representation of local coordinates of (a) model A and (b) model B. Grey areas are zirconia grains.

**Fig. 4** Magnitude of the average stress of $Al_2O_3$ and $ZrO_2$ grains after cooling. (a) Model A and (b) model B. $\sigma_x$ and $\sigma_y$ denote normal stresses in x and y directions shown in Fig.4, respectively. $\tau_{xy}$ is shear stress.

**Fig. 5** Crack density versus applied normal stress chart. Shear stresses were applied with the normal stresses together.

**Fig. 6** Illustrations of crack distribution in model A with respect to magnitude of applied contact stress: (a) a normal (shear) stress of 100 MPa (40 MPa), (b) a normal (shear) stress of 500 MPa (200 MPa), (c) a normal (shear) stress of 100 MPa (40 MPa) in the model with a void, and (d) a normal (shear) stress of 500 MPa (200 MPa) in model with a void. Grey areas are $Al_2O_3$ and dark ones are $ZrO_2$. Dark lines between grains denote cohesive layers and white ones indicate cracks.

**Fig. 7** Illustrations of crack distribution in model B with respect to applied contact stress: (a) a normal (shear) stress of 100 MPa (40 MPa), and (b) a normal (shear) stress of 500 MPa (200 MPa), (c) a normal (shear) stress of 100 MPa (40 MPa) in the model with a void, and (d) a normal (shear) stress of 500 MPa (200 MPa) in model with a void. Grey areas are $Al_2O_3$, and dark ones are $ZrO_2$. Dark lines between grains denote cohesive layers, and white ones indicate cracks.

**Fig. 8** Illustrations of crack distribution in model A with a void according to number of fatigue cycles: (a) cycle 1, (b) cycle 5, and (c) cycle 10. Grey areas are $Al_2O_3$ and dark ones are $ZrO_2$. Dark lines between grains denote cohesive layers and white ones indicate cracks. Repeated loading was applied, ranging from zero to 100 MPa for normal stress (40 MPa for shear stress).

**Fig. 9** Illustrations of crack distribution in model B with a void with respect to number of fatigue cycles: (a) cycle 1, (b) cycle 3, and (c) cycle 5. Grey areas are $Al_2O_3$, and dark ones are $ZrO_2$. Dark lines between grains denote cohesive layers and white ones indicate cracks. Repeated loading was applied, ranging from zero to 100 MPa for normal stress (40 MPa for shear stress).

**Fig. 10** Evolution of crack density with respect to number of fatigue cycles. Markers denote calculated data. Continuous lines denote curve fits.

**Figures**

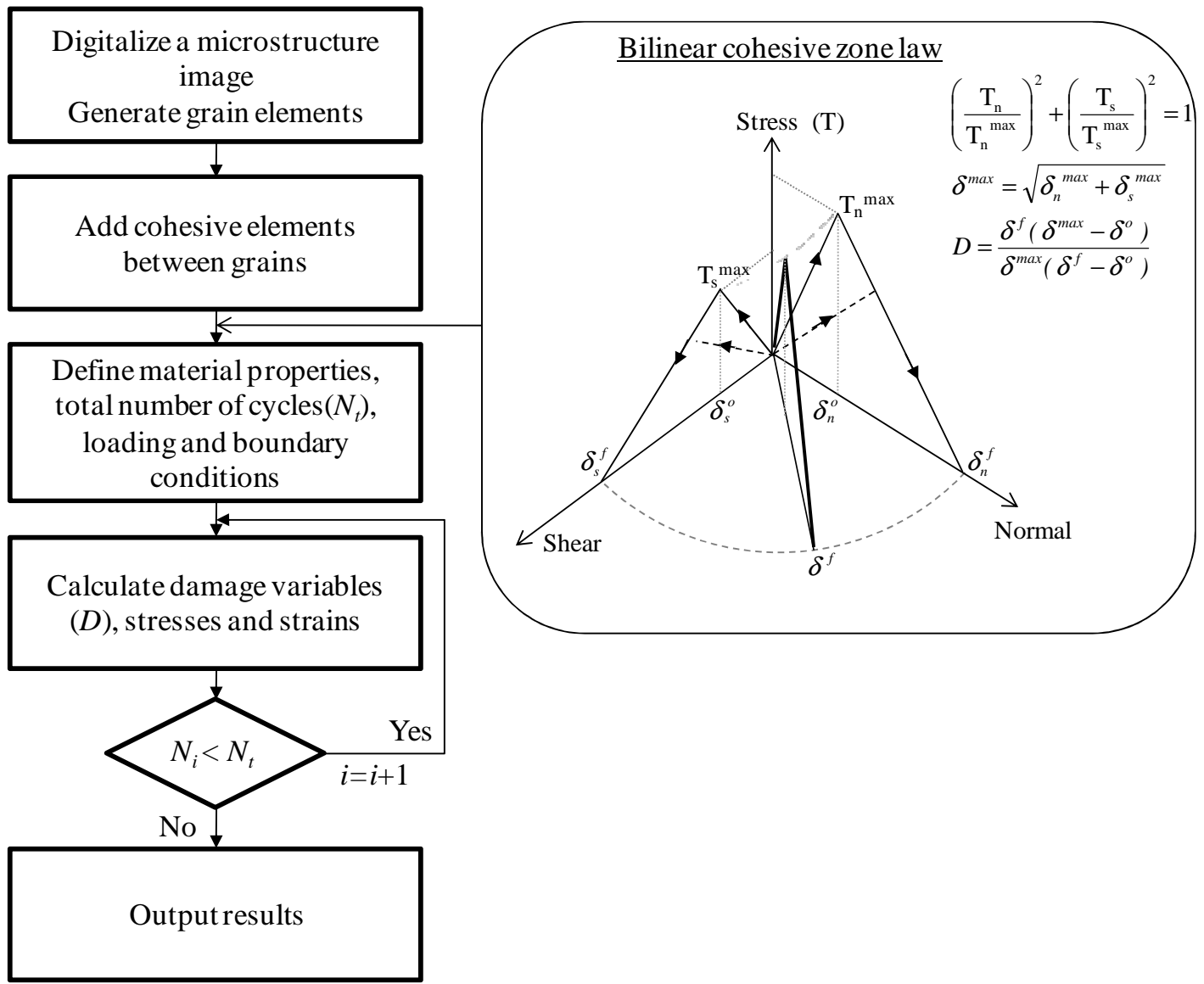

**Fig. 1** Simulation algorithm for determining damage. $N_i$ denotes the $i^{th}$ cycle.

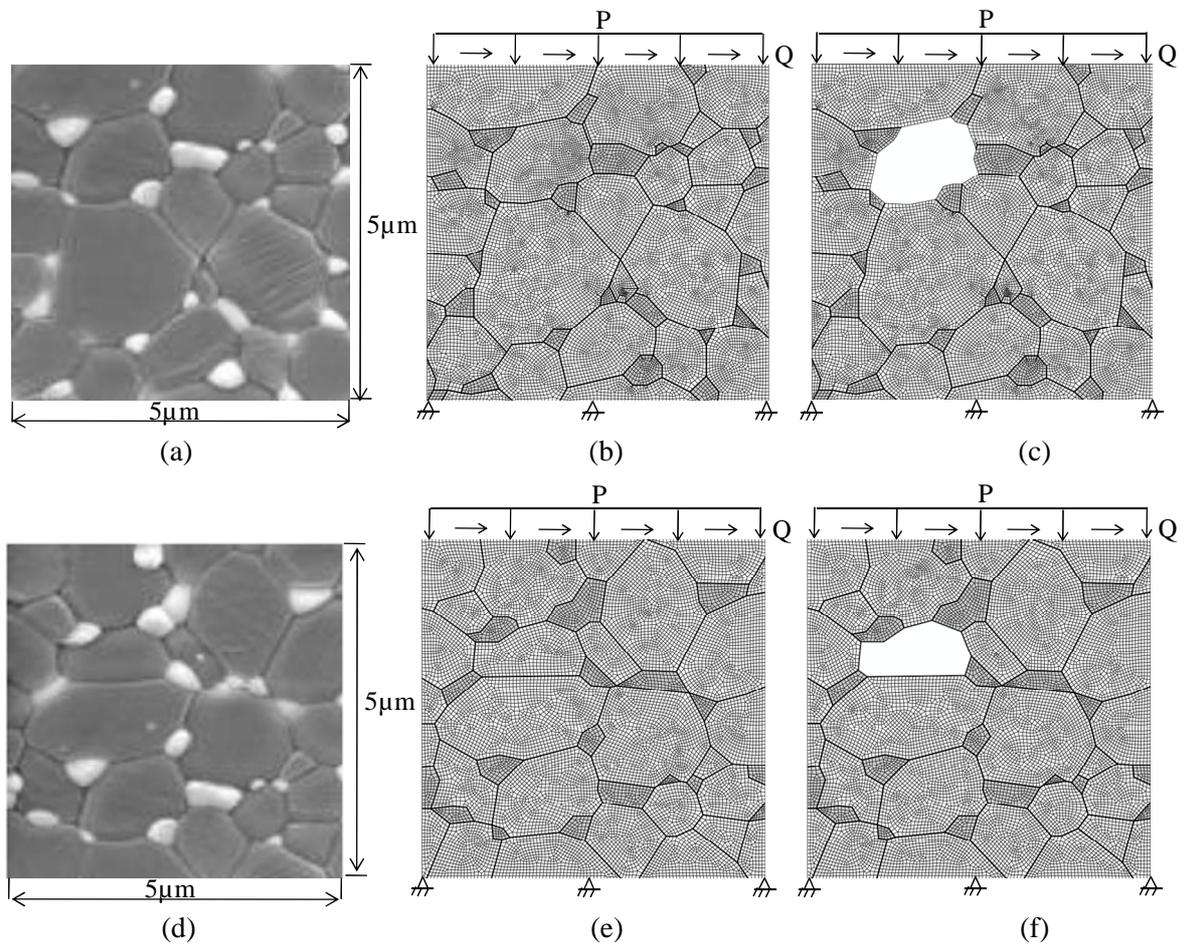

**Fig. 2** Images of $Al_2O_3$-10%$ZrO_2$ microstructures [14] and finite element models. The middle images (b, e) are models generated with the actual images (a, d). For convenience, (b) and (e) are denoted as model A and model B, respectively. A single alumina grain in each model was deleted for investigating the effect of a void (c, f).

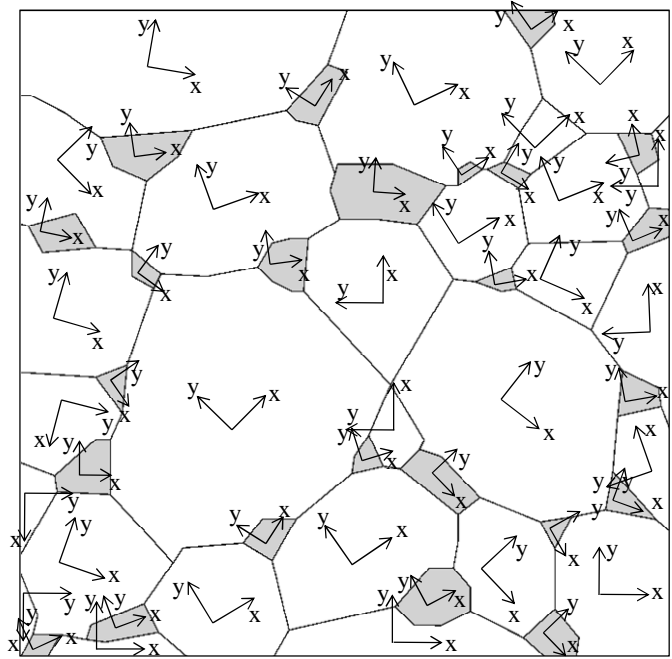

(a)

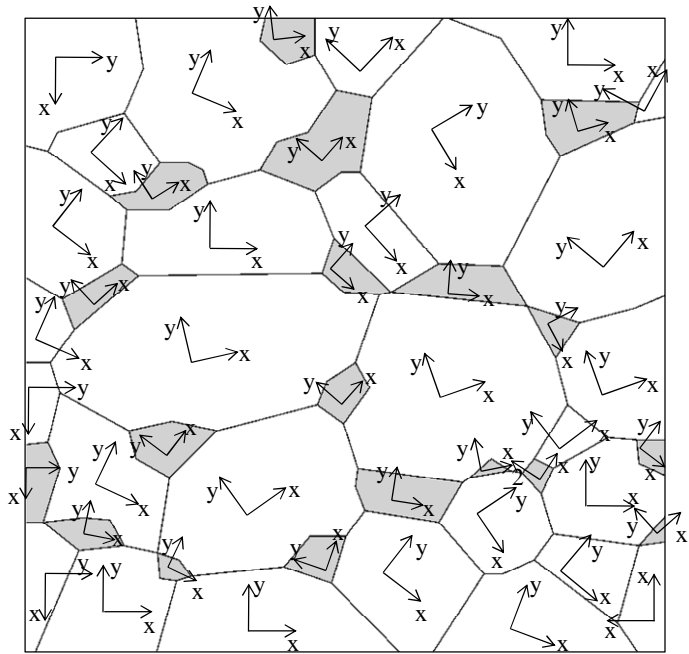

(b)

**Fig. 3** Schematic representation of local coordinates of (a) model A and (b) model B. Grey areas are zirconia grains.

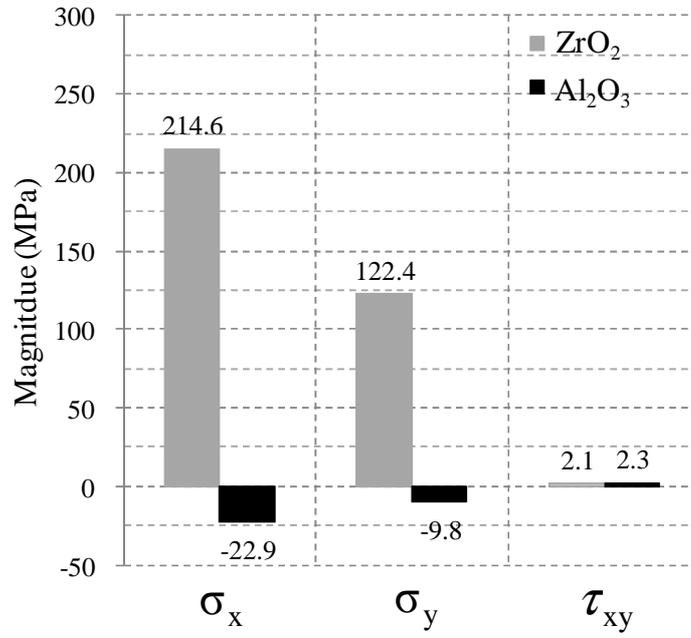

(a)

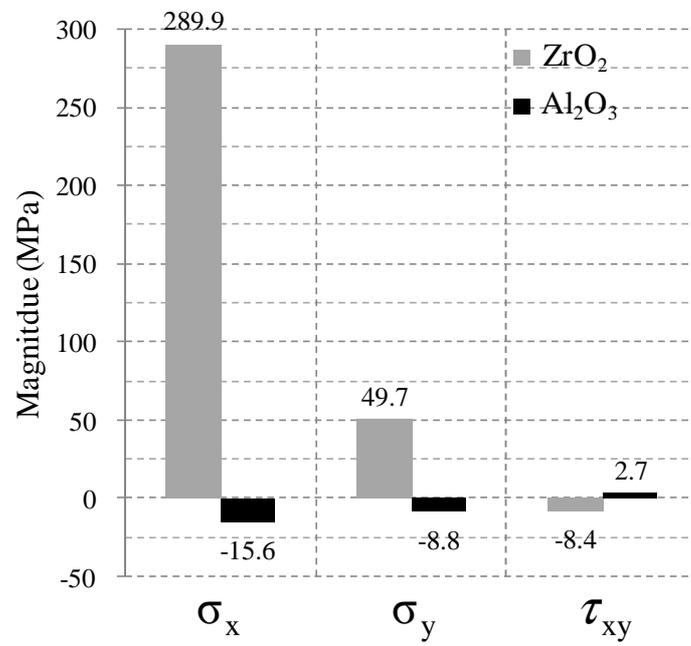

(b)

**Fig. 4** Magnitude of the average stress of $Al_2O_3$ and $ZrO_2$ grains after cooling. (a) Model A and (b) model B. $\sigma_x$ and $\sigma_y$ denote normal stresses in x and y directions shown in Fig.4, respectively. $\tau_{xy}$ is shear stress.

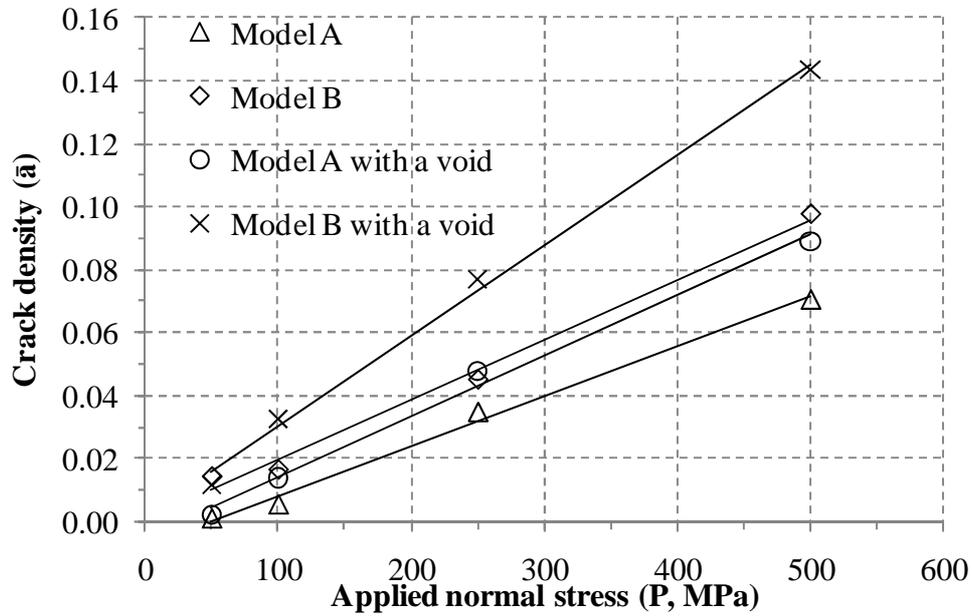

**Fig. 5** Crack density versus applied normal stress chart. Shear stresses were applied with the normal stresses together.

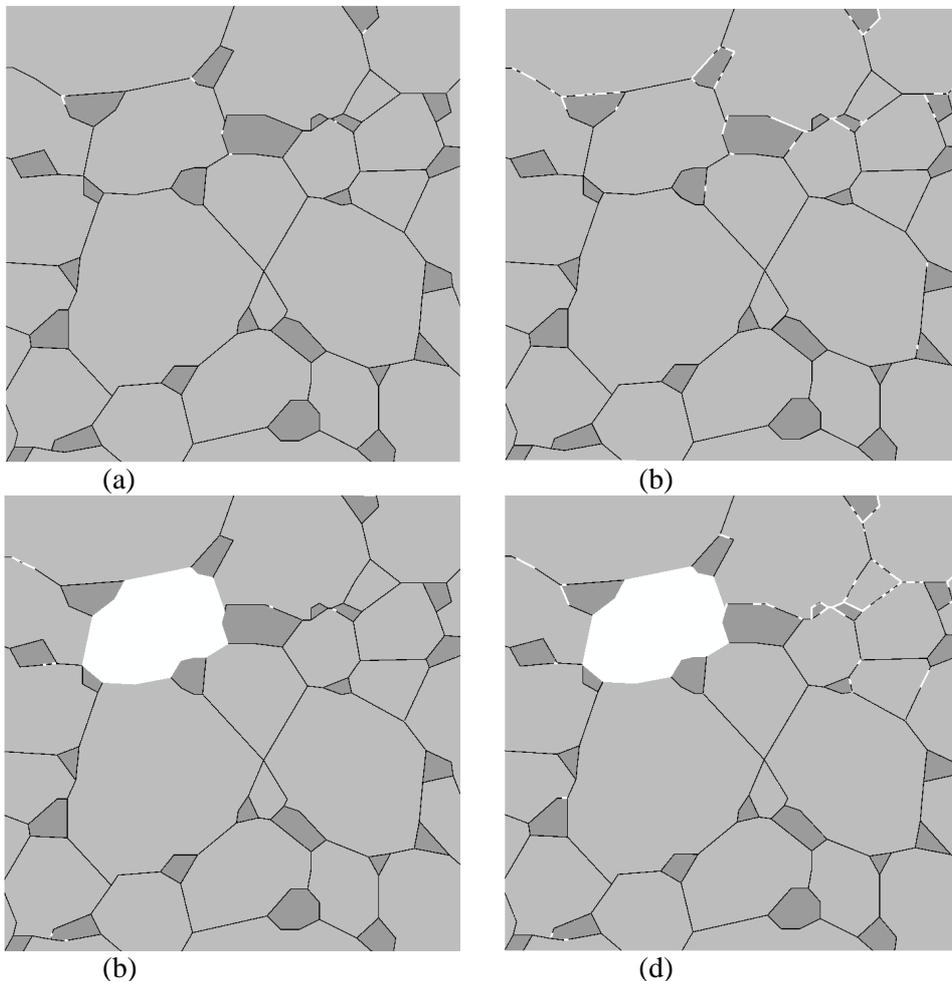

**Fig. 6** Illustrations of crack distribution in model A with respect to magnitude of applied contact stress: (a) a normal (shear) stress of 100 MPa (40 MPa), (b) a normal (shear) stress of 500 MPa (200 MPa), (c) a normal (shear) stress of 100 MPa (40 MPa) in the model with a void, and (d) a normal (shear) stress of 500 MPa (200 MPa) in model with a void. Grey areas are $Al_2O_3$ and dark ones are $ZrO_2$. Dark lines between grains denote cohesive layers and white ones indicate cracks.

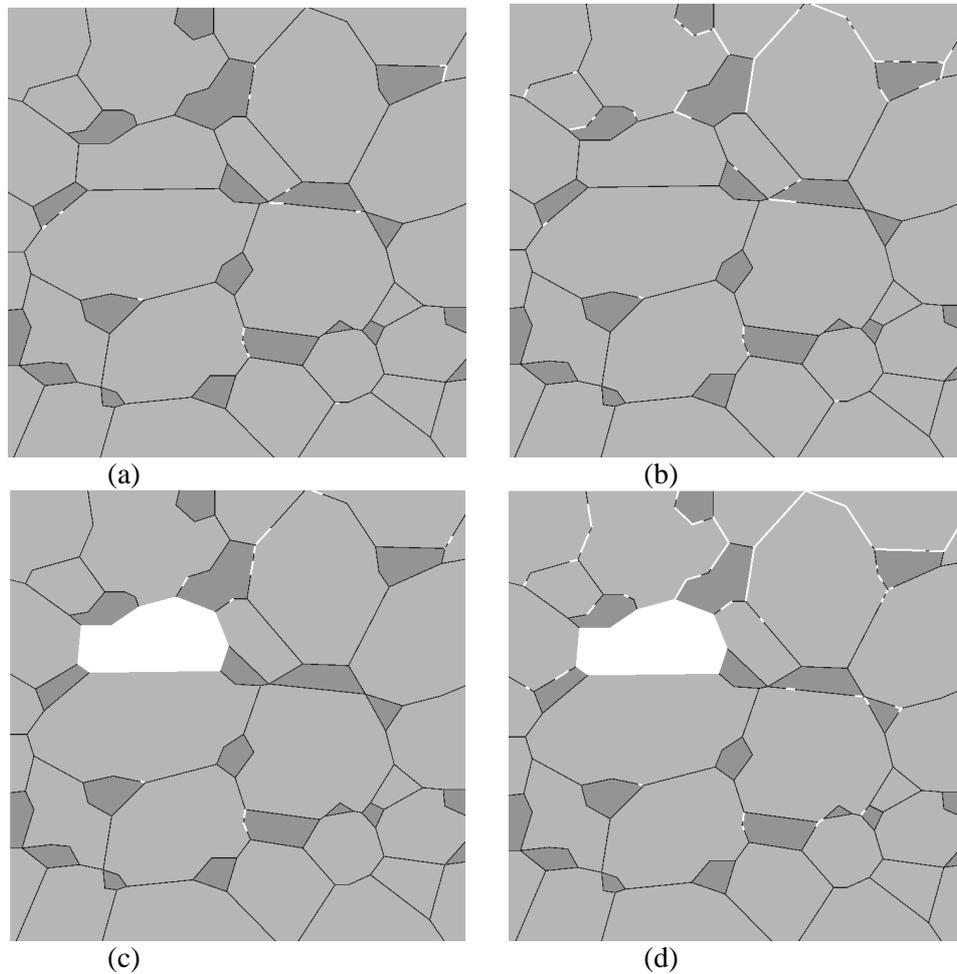

**Fig. 7** Illustrations of crack distribution in model B with respect to applied contact stress: (a) a normal (shear) stress of 100 MPa (40 MPa), and (b) a normal (shear) stress of 500 MPa (200 MPa), (c) a normal (shear) stress of 100 MPa (40 MPa) in the model with a void, and (d) a normal (shear) stress of 500 MPa (200 MPa) in model with a void. Grey areas are $Al_2O_3$, and dark ones are $ZrO_2$. Dark lines between grains denote cohesive layers, and white ones indicate cracks.

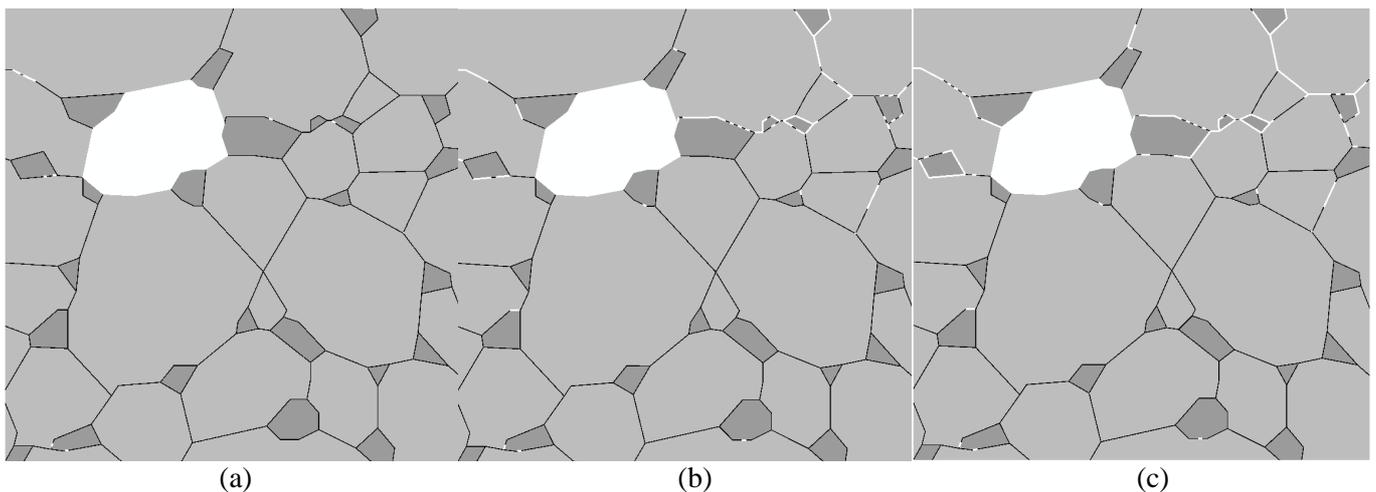

**Fig. 8** Illustrations of crack distribution in model A with a void according to number of fatigue cycles: (a) cycle 1, (b) cycle 5, and (c) cycle 10. Grey areas are $Al_2O_3$ and dark ones are $ZrO_2$. Dark lines between grains denote cohesive layers and white ones indicate cracks. Repeated loading was applied, ranging from zero to 100 MPa for normal stress (40 MPa for shear stress).

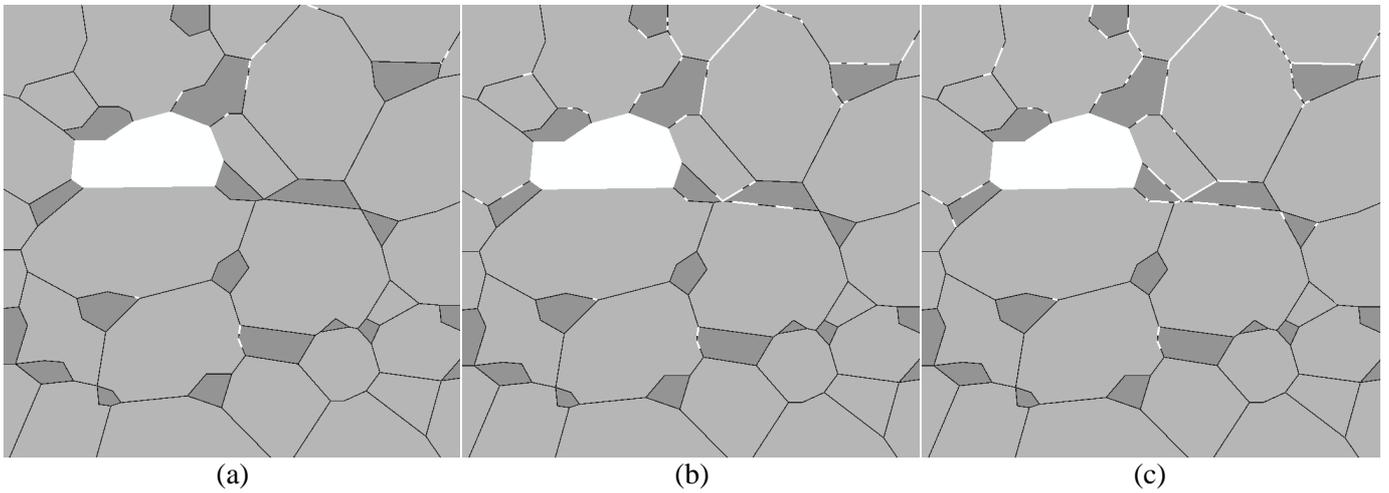

**Fig. 9** Illustrations of crack distribution in model B with a void with respect to number of fatigue cycles: (a) cycle 1, (b) cycle 3, and (c) cycle 5. Grey areas are $Al_2O_3$, and dark ones are $ZrO_2$. Dark lines between grains denote cohesive layers and white ones indicate cracks. Repeated loading was applied, ranging from zero to 100 MPa for normal stress (40 MPa for shear stress).

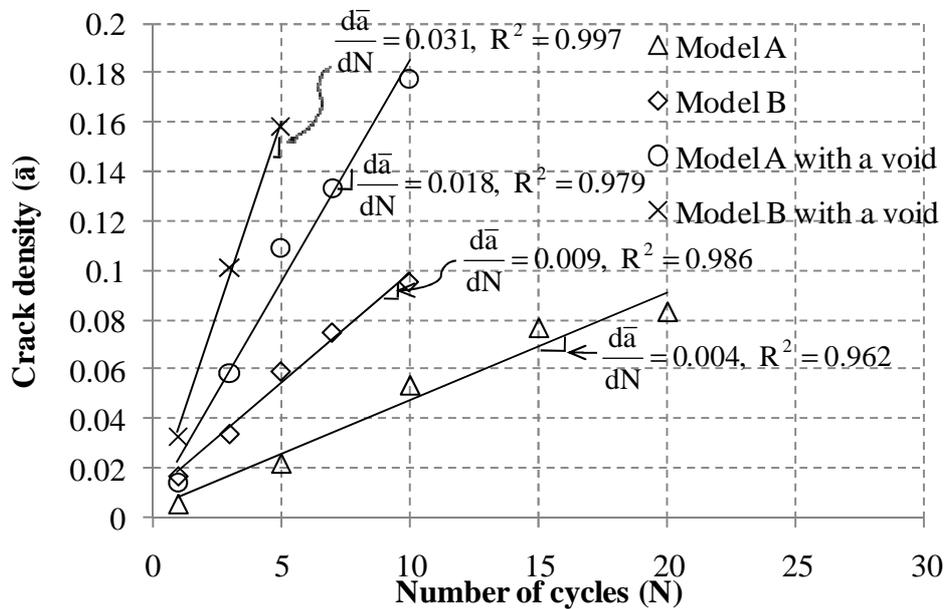

**Fig. 10** Evolution of crack density with respect to number of fatigue cycles. Markers denote calculated data. Continuous lines denote curve fits.